# NON-ASYMPTOTIC INFERENCE IN INSTRUMENTAL VARIABLES ESTIMATION

by

Joel L. Horowitz
Department of Economics
Northwestern University
Evanston, IL 60208

September 2018

## ABSTRACT

This paper presents a simple method for carrying out inference in a wide variety of possibly nonlinear IV models under weak assumptions. The method is non-asymptotic in the sense that it provides a finite sample bound on the difference between the true and nominal probabilities of rejecting a correct null hypothesis. The method is a non-Studentized version of the Anderson-Rubin test but is motivated and analyzed differently. In contrast to the conventional Anderson-Rubin test, the method proposed here does not require restrictive distributional assumptions, linearity of the estimated model, or simultaneous equations. Nor does it require knowledge of whether the instruments are strong or weak. It does not require testing or estimating the strength of the instruments. The method can be applied to quantile IV models that may be nonlinear and can be used to test a parametric IV model against a nonparametric alternative. The results presented here hold in finite samples, regardless of the strength of the instruments.

Key Words: Weak instruments, normal approximation, finite-sample bounds

JEL Listing: C21, C26

___________________________________________________________________________________
I thank Ivan Canay, Xu Cheng, Denis Chetverikov, Whitney Newey, and Vladimir Spokoiny for helpful comments and discussions, and Caleb Kwon for research assistance.

# NON-ASYMPTOTIC INFERENCE IN INSTRUMENTAL VARIABLES ESTIMATION

## 1. INTRODUCTION

Instrumental variables (IV) estimation is an important and widely used method in applied econometrics. However, inference based on IV estimates is problematic if the instruments are weak or the number of instruments is large. With weak or many instruments, conventional asymptotic approximations can be highly inaccurate. Nelson and Startz (1990a, 1990b) illustrate this problem with a simple model. Angrist and Krueger (1991) is a well-known empirical application in which the problem arises. Bound, Jaeger, and Baker (1995) and Hansen, Hausman, and Newey (2008) provide detailed discussions of the problems of inference in Angrist and Krueger (1991).

Exact finite sample methods for inference in IV estimation exist but depend on strong assumptions about the population from which the data are sampled and/or require the model being estimated to be linear in the unknown parameters. This paper presents a simple method for carrying out inference in IV models that is easy to implement and does not rely on strong assumptions or asymptotic approximations. The method is a modification of the well-known Anderson-Rubin (1949) test but does not require restrictive distributional assumptions, linearity of the estimated model, or knowledge of whether the instruments are strong or weak. It does not require testing or estimating the strength of the instruments. The results presented here hold in finite samples under mild assumptions that are easy to understand, regardless of the strength of the instruments. The method described here also can be used to carry out inference in quantile IV models that may be nonlinear and to test a parametric IV model or quantile IV model against a nonparametric alternative.

There is a long history of research aimed at developing reliable methods for inference in IV estimation, and the associated literature is very large. One stream of research has been concerned with deriving the exact finite-sample distributions of IV estimators and test statistics based on IV estimators. The test of Anderson and Rubin (1949) is a well-known early example of this research. Phillips (1983) and the references therein present additional results of early research in this stream. Recent examples of exact finite-sample results include Andrews and Marmer (2008); Andrews, Moreira, and Stock (2006); Dufour and Taamouti (2005); and Moreira (2003, 2009). Obtaining exact finite-sample results often requires strong assumptions about the population from which the data are sampled. Most results are based on the assumption that the data are generated by a linear simultaneous equations model whose stochastic disturbances are homoskedastic and normally distributed with a known covariance matrix. Andrews and Marmer (2008) assume a linear model but not a system of simultaneous equations or normality.

Another stream of research derives non-standard or higher order asymptotic approximations to the distributions of IV estimators and test statistics. Staiger and Stock (1997), Wang and Zivot (1998), Stock and Wright (2000), Andrews and Cheng (2012), Andrews and Mikusheva (2016), and Carrasco and Tchuente (2016) are examples of the literature on non-standard first-order asymptotic approximations. Examples of higher-order expansions include Holly and Phillips (1979), Rothenberg (1984), and the references therein. Kitamura and Stutzer (1997); Imbens, Spady, and Johnson (1998); Newey and Smith (2004); and Guggenberger and Smith (2005), among others, discuss estimators with improved higher-order properties.

A third stream of research aims at deriving the asymptotic distributions of estimators and test statistics when the number of instruments is an increasing function of the sample size and, with most methods, the instruments may be weak. Andrews and Stock (2007a) review much of this literature. Examples include Bekker (1994); Kleibergen (2002); Andrews and Stock (2007b); Hansen, Hausman, and Newey (2008); and Newey and Windmeijer (2009). Some research in this stream includes weakening the assumptions used to obtain the exact finite-sample distributions of certain statistics and finding the resulting asymptotic distributions of these statistics. See, for example, Andrews, Moreira, and Stock (2006) and Andrews and Soares (2007).

The approach taken here is different from the approaches in the foregoing literature. A hypothesis $H_0$ about a finite-dimensional parameter can be tested by using a test statistic that is a quadratic form in the sample analog of the identifying moment conditions. This statistic is a non-Studentized version of the Anderson-Rubin (1949) statistic (see, also, the $S$ statistic of Stock and Wright 2000) but is motivated and analyzed differently. It can also be interpreted as a statistic for testing the hypothesis that a multivariate mean is zero. Except in special cases, the finite-sample distribution of the statistic is a complicated function of the unknown population distribution of the observed variables. We overcome this problem by approximating the unknown population distribution with a normal distribution. The finite-sample distribution of the resulting approximate test statistic can be computed by simulation with any desired accuracy. We obtain a finite-sample bound on the difference between the true and nominal probabilities of rejecting a correct $H_0$ (the error in the rejection probability or ERP) when the critical value is obtained by using the simulation procedure. In contrast to the tests cited in the foregoing two paragraphs, the test presented here is non-asymptotic in the sense that it provides a finite-sample bound on the ERP

Advantages of the method include:

1. It is easy to understand and implement.

2. It applies to a wide variety of linear and nonlinear models, including mean and quantile IV models.

3. It can be used to test a parametric mean or quantile IV model against a nonparametric alternative.

4. It does not require identification of the parameter about which inference is made.

5. It does not require strong distributional assumptions, simultaneous equations, knowledge of whether the instruments are weak, or testing for weakness.

6. It provides an exact finite-sample bound on the difference between the true and nominal probabilities of rejecting a correct null hypothesis.

Many other methods have some of these features, but we are not aware of another method that has all of them. The method described here does not provide a test whose exact finite-sample size is known or that has the optimality properties of some other tests.

The normal approximation used here is a multivariate generalization of the Berry-Esséen theorem and due to Bentkus (2003). Other normal approximations have been developed by Chernozhukov, Chetverikov, and Kato (2017) and Spokoiny and Zhilova (2015), among many others. Chernozhukov, Chetverikov, and Kato (2013) and Spokoiny and Zhilova (2015) provide reviews. The error of Bentkus's (2003) approximation converges to zero more rapidly as the sample size increases than errors of the other approximations when the number of instruments and exogenous covariates is small compared to the sample size.

Section 2 of this paper describes the version of the standard IV model that we consider, the hypotheses that are tested, and the test method. Section 3 presents the main result for the model of Section 2. Section 4 presents extensions to quantile IV models and to testing a parametric model against a nonparametric alternative. Section 5 presents the results of a Monte Carlo investigation of the numerical performance of the method. Section 6 presents two empirical applications, and Section 7 presents conclusions. The proofs of theorems are presented in the appendix, which is Section 8.

## 2. THE STANDARD IV MODEL, HYPOTHESES, AND METHOD

### 2.1 *The Model and Hypotheses*

The model considered in this this section and Section 3 is

(2.1) $\quad Y = g(X,\theta) + U; \quad E(U \mid Z) = 0$,

where $Y$ is a scalar outcome variable, $X$ is a vector of covariates, $U$ is a scalar random variable, $g$ is a known real-valued function, and $\theta$ is an unknown finite-dimensional vector of constant parameters. The parameter $\theta$ is contained in a parameter set $\Theta \subset \mathbb{R}^d$ for some $d \geq 1$. One or more components of $X$ may be endogenous. $Z$ is a vector of instruments for $X$. The elements of $Z$ include any exogenous components of $X$. $U$ can have any (possibly unknown) form of heteroskedasticity that is consistent with (2.1) and the regularity conditions given in Section 3. Let $q$ denote the dimension of $Z$. The dimension of $X$ does not enter the notation used in this paper.

Let $\{Y_i, X_i, Z_i : i = 1,...,n\}$ be an independent random sample from the distribution of $(Y, X, Z)$. Let $Z_{ij}$ ($i = 1,...,n; j = 1,...,q\}$ denote the $j$'th component of $Z_i$. For any $\theta \in \Theta$, define

$$T_n(\theta) = n^{-1} \sum_{j=1}^{q} \left\{ \sum_{i=1}^{n} Z_{ij}[Y_i - g(X_i, \theta)] \right\}^2 .$$

Denote the covariance matrix of the random vector $Z[Y - g(X, \theta)]$ by $\Sigma(\theta)$.

We consider two hypotheses about $\theta$, one simple and one composite. The simple null hypothesis is

(2.2) $\quad H_0 : \theta = \theta_0$

for some $\theta_0 \in \Theta$ against the alternative

$H_1 : \theta \neq \theta_0$.

Under hypothesis (2.2), $\Sigma(\theta_0) = E(ZZ'U^2)$. The matrix $E(ZZ'U^2)$ will be denoted by $\Sigma$ without the argument $\theta_0$ when this will not cause confusion.

To describe the composite null hypothesis, let $\vartheta$ be a subvector of $\theta$, and let $\theta = (\vartheta', \beta')'$. The composite null hypothesis is

(2.3) $\quad H_0 : \vartheta = \vartheta_0$.

The alternative hypothesis is

$H_1 : \vartheta \neq \vartheta_0$.

For the composite hypothesis, define $\mathcal{B} = \{b : (\vartheta_0', b')' \in \Theta\}$ and $\hat{\theta} = \arg\min\{T_n(\theta) : \theta \in \mathcal{B}\}$. A hypothesis about a linear combination of components of $\theta$ can be put into the form (2.2) or (2.3) by redefining the components of $\theta$ and, therefore, does not require a separate formulation.

### 2.2 Test Statistics

The statistic for testing the simple null hypothesis (2.2) is $T_n(\theta_0)$. Let $c_\alpha(\theta_0)$ denote the $\alpha$-level critical value for testing the simple hypothesis $H_0 : \theta = \theta_0$. That is, $c_\alpha(\theta_0)$ is the $1-\alpha$ quantile of the distribution of $T_n(\theta_0)$. The test of the composite null hypothesis (2.3) consists of testing whether there is a $b \in \mathcal{B}$ for which the point $(\vartheta_0', b')'$ is contained in a confidence region for $\theta$. Therefore, testing (2.3) can be reduced to testing (2.2). Define $\breve{\theta}(b) = (\vartheta_0', b')'$ for any $b \in \mathcal{B}$. Let $c_\alpha(b)$ denote the $\alpha$-level critical value for testing the simple hypothesis $H_0 : \theta = \breve{\theta}(b)$. That is, $c_\alpha(b)$ is the $1-\alpha$ quantile of the distribution of $T_n[\breve{\theta}(b)]$. If hypothesis (2.3) is correct, then the simple hypothesis $H_0: \theta = \breve{\theta}(\beta_0)$ is correct for some $\beta_0 \in \mathcal{B}$.

The critical values $c_\alpha(\theta_0)$ and $c_\alpha(b)$ are unknown in applications. Let $\hat{c}_\alpha(\theta_0)$ and $\hat{c}_\alpha(b)$, respectively, denote the estimators of these quantities described in Section. 2.3.  Hypothesis (2.2) is rejected at the $\alpha$ level if $T_n(\theta_0) > \hat{c}_\alpha(\theta_0)$. Hypothesis (2.3) is rejected at the $\alpha$ level if $T_n[\breve{\theta}(b)] > \hat{c}_\alpha(b)$ for every $b \in \mathcal{B}$. Computationally, the test of (2.3) consists of solving the nonlinear optimization problem

(2.4) $\quad \underset{b \in \mathcal{B}}{\text{minimize}} : \{T_n[\breve{\theta}(b)] - \hat{c}_\alpha(b)\}$.

Hypothesis (2.3) is rejected if the optimal value of the objective function in (2.4) exceeds zero. Under hypothesis (2.3), the rejection probability does not exceed $P[T_n(\theta_0) > \hat{c}_\alpha(\theta_0)]$, where $\theta_0$ is the true value of $\theta$ in (2.1). We obtain an upper bound on $P[T_n(\theta_0) > \hat{c}_\alpha(\theta_0)]$ that does not depend on $\theta_0$. Therefore, it suffices to bound the probability of rejecting hypothesis (2.2).

In applications, the set $\mathcal{B}$ in (2.4) can be replaced by a much smaller set or even a single point. Let $\hat{\mathcal{B}}$ be an arbitrarily small neighborhood of $\hat{\theta}$. If $H_0$ is true, then $P(\theta_o \in \hat{\mathcal{B}}) \to 1$ at an exponentially fast rate as $n \to \infty$. Consequently, $\mathcal{B}$ in (2.4) can be replaced by an arbitrarily small neighborhood $\hat{\mathcal{B}}$. Moreover, the optimal value of the objective function of (2.4) is typically very close to $T_n(\hat{\theta}) - \hat{c}_\alpha(\hat{\theta})$. Therefore, $\breve{\theta}(b)$ can be replaced by $\hat{\theta}$ and $T_n[\breve{\theta}(b)] - \hat{c}_\alpha(b)$ with $T_n(\hat{\theta}) - \hat{c}_\alpha(\hat{\theta})$ in applications except, possibly, if $T_n(\hat{\theta}) - \hat{c}_\alpha(\hat{\theta})$ is close to 0. Solving the nonlinear optimization problem (2.4) is unnecessary if $T_n(\hat{\theta}) - \hat{c}_\alpha(\hat{\theta})$ is used in place of $T_n[\breve{\theta}(b)] - \hat{c}_\alpha(b)$.

The $\alpha$ level test based on $T_n(\theta_0)$ has asymptotic power exceeding $\alpha$ against alternatives whose "distance" from $H_0$ is $O(n^{-1/2})$, but the test does not have optimal asymptotic power in general. The statistic $T_n(\theta_0)$ and its quantile analog that is described in Section 4 are designed to avoid the need for estimating $\theta$ and the inverses of matrices that may be nearly singular. Estimators of $\theta$ and inverses of nearly singular matrices can be very imprecise, and non-asymptotic inference about an estimator of $\theta$ is difficult or impossible in nonlinear models. A test that requires possibly imprecise estimation of $\theta$ and inverses of matrices can have low finite-sample power, and there can be a large difference between the true and nominal probabilities with which the test rejects a correct null hypothesis.

2.3 *The Test Procedure*

Under the simple null hypothesis (2.2),

(2.5) $$T_n(\theta_0) = n^{-1} \sum_{j=1}^{q} \left( \sum_{i=1}^{n} Z_{ij} U_i \right)^2,$$

where $U_i = Y_i - g(X_i, \theta_0)$. If the distribution of $ZU$ were known, the finite-sample distribution of $T_n(\theta_0)$ could be computed from (2.5) by simulation. However, the distribution of $ZU$ is unknown. To overcome this problem, define $V$ to be the $q \times 1$ vector whose $j$'th component ($j = 1, ..., q$) is

$$V_j = n^{-1/2} \sum_{i=1}^{n} Z_{ij} U_i.$$

Then $E(V) = 0$, $E(VV') = \Sigma$, and $T_n(\theta_0) = V'V$. Let $\hat{\Sigma}$ be a consistent estimator of $\Sigma$, and let $\hat{V}$ be a $q \times 1$ random vector that is distributed as $N(0, \hat{\Sigma})$. Define

(2.6) $\hat{T}_n(\theta_0) = \hat{V}'\hat{V}$.

The distribution of $\hat{T}_n(\theta_0)$ can be computed with any desired accuracy by simulation. Let $\hat{c}_\alpha(\theta_0)$ denote the $1 - \alpha$ quantile of the distribution of $\hat{T}_n(\theta_0)$. Then

(2.7) $P[\hat{T}_n(\theta_0) > \hat{c}_a(\theta_o)] = \alpha$.

Section 3 presents a non-asymptotic upper bound on $|P[T_n(\theta_0) > \hat{c}_\alpha(\theta_0)] - \alpha|$ that holds with high probability under $H_0$. Accordingly, the test procedure proposed here consists of:

1. Estimate $\Sigma$ using the estimator $\hat{\Sigma}$ consisting of the $q \times q$ matrix whose $(j, k)$ component is

$$\hat{\Sigma}_{jk} = n^{-1} \sum_{i=1}^{n} Z_{ij} Z_{ik} [Y_i - g(X_i, \theta_0)]^2 - \hat{\mu}_j \hat{\mu}_k,$$

where

$$\hat{\mu}_j = n^{-1} \sum_{i=1}^{n} Z_{ij} [Y_i - g(X_i, \theta_0)].$$

2. Use simulation to compute the distribution of $\hat{T}_n(\theta_0)$ and the critical value $\hat{c}_\alpha(\theta_0)$ by repeatedly drawing $\hat{V}$ from the $N(0, \hat{\Sigma})$ distribution.

3. Reject $H_0$ at the $\alpha$ level if $T_n(\theta_0) > \hat{c}_\alpha(\theta_0)$.

The critical value of $\hat{T}_n[\breve{\theta}(b)]$, $\hat{c}_\alpha(b)$, is estimated by replacing $\theta_0$ with $\breve{\theta}(b)$ in steps 1-2. Section 5 presents Monte Carlo evidence on the numerical performance of this procedure.

It is not difficult to derive the asymptotic distribution of $T_n(\theta_0)$. See Theorem 3.2 in Section 3. This distribution depends on the unknown population parameter $\Sigma$. The finite-sample distribution of $\hat{T}_n(\theta_0)$ is the asymptotic distribution of $T_n(\theta_0)$ with $\Sigma$ replaced by $\hat{\Sigma}$. Thus, the foregoing computational procedure is a simulation method to compute the estimated asymptotic distribution of $T_n(\theta_0)$. The main result for model (2.1), which is given in Theorem 3.1, is a bound on the difference between the unknown finite-sample distribution of $T_n(\theta_0)$ and its estimated asymptotic distribution, which is the finite-sample distribution of $\hat{T}_n(\theta_0)$. A similar result for the quantile version of $T_n(\theta_0)$ is given in Theorem 4.1. The distributions of $T_n(\theta_0)$, $\hat{T}_n(\theta_0)$, and their quantile versions are not chi-square because, to avoid the need for inverting estimated matrices, these statistics are not Studentized.

## 3. MAIN RESULT FOR MODEL (2.1)

This section presents the non-asymptotic upper bound on $|P[T_n(\theta_0) > \hat{c}_\alpha(\theta_0)] - \alpha|$ in model (2.1). Make the following assumptions, which are stated in a way that accommodates tests of both simple hypothesis (2.2) and composite hypothesis (2.3).

<u>Assumption 1</u>: $\{Y_i, X_i, Z_i : i = 1,...,n\}$ is an independent random sample from the distribution of $(Y, X, Z)$. (ii) $\theta \in \Theta \in \mathbb{R}^d$.

<u>Assumption 2</u>: (i) $\Sigma(\theta)$ is nonsingular for every $\theta \in \Theta$. (ii) Let $\Sigma_{jk}^{-1}(\theta)$ denote the $(j,k)$ component of $\Sigma^{-1}(\theta)$. There is a constant $C_\Sigma < \infty$ such that $|\Sigma_{jk}^{-1}(\theta)| \leq C_\Sigma$ for each $j, k = 1,...,q$ and every $\theta \in \Theta$.

Define the $q \times 1$ vectors $\xi = ZU$ and $\zeta = \Sigma^{-1/2}\xi$. Define the $q \times q$ matrix $\eta = ZZ'U^2$. Let $\xi_j$ and $\zeta_j$ ($j = 1,...,q$) denote the $j$'th components of $\xi$ and $\zeta$, respectively. Let $\eta_{jk}$ ($j, k = 1,...,q$) denote the $(j,k)$ component of $\eta$.

<u>Assumption 3</u>: (i) There is a finite constant $m_3$ such that $E|\zeta_j|^3 \leq m_3$ for every $j = 1,...,q$. (ii) There is a finite constant $\ell \geq \max[\max_j E(\xi_j^2), \max_{j,k} E(\eta_{jk}^2)]$ such that $E|\xi_j|^r \leq \ell^{r-1} r!$ and $E|\eta_{jk}|^r \leq \ell^{r-1} r!$ for every $r = 3, 4, 5,...$ and $j, k = 1,...,q$.

Assumption 1 specifies the sampling process. Assumption 2 establishes mild non-singularity conditions. For example, if $U$ and $Z$ are independent, then Assumption 2 requires $\text{cov}(Z)$ to be non-singular. Assumption 3 requires the distributions of the components of $\xi$ and $\eta$ to be thin-tailed. The assumption is satisfied, for example, if these distributions are sub-exponential. It is needed to establish the conditions of certain probability inequalities that are used in the proof of Theorem 3.1.

For any $t > 0$ define

$$r(t) = \left(\frac{6\ell t}{n}\right)^{1/2}$$

and

$$\tilde{r}(t) = C_\Sigma q^2 [r(t) + r(t)^2].$$

The following theorem gives the non-asymptotic upper bound on $|P[T_n(\theta_0) > \hat{c}_\alpha(\theta_0)] - \alpha|$ in model (2.1). The theorem is stated in terms of a test of hypothesis (2.2). As was explained in Section 2.2, testing hypothesis (2.3) can be reduced to testing hypothesis (2.2).

<u>Theorem 3.1</u>: Let assumptions 1-3 and hypothesis (2.2) hold. Define $\hat{c}_\alpha(\theta_0)$ as in (2.7), but treat it as a non-stochastic constant in the following inequality. If $\max[q\tilde{r}(t), r(t)] < 1$, then

$$(3.1) \quad |P[T_n(\theta_0) > \hat{c}_\alpha(\theta_0)] - \alpha| \le \frac{400 q^{7/4} m_3}{n^{1/2}} + \min\begin{cases} q 2^{q+1} \tilde{r}(t - 2\log q) \\ \frac{1}{\sqrt{2}} \{\tilde{r}(t - 2\log q) - \log[1 - \tilde{r}(t - 2\log q)]\}^{1/2} \end{cases}$$

with probability at least $1 - 4e^{-t}$. ∎

The critical value $\hat{c}_\alpha(\theta_0)$ is a function of the random matrix $\hat{\Sigma}$ and, therefore, a random variable. However, in the probability expression on the left-hand side of (3.1), $\hat{c}_\alpha(\theta_0)$ is treated as a non-stochastic constant. When $\hat{c}_\alpha(\theta_0)$ is treated this way, inequality (3.1) holds only if $\hat{\Sigma}$ satisfies conditions (8.2) and (8.3) in the appendix. If these conditions are not satisfied, then (3.1) may not hold. The probability that the conditions are satisfied is at least $1 - 4e^{-t}$.

The right-hand side of (3.1) does not depend on how $X$ is related to the instruments. In particular, the upper bound on the probability of rejecting a correct simple or composite null hypothesis does not depend on the strength or weakness of the instruments.

The non-asymptotic bound in (3.1), like other large deviations bounds in statistics and the Berry-Esséen bound, tends to be loose unless $n$ is large because it accommodates "worst case" distributions of $(Y, X, Z)$. For example, the distribution of $Z[Y - g(X_i, \theta_0)]$ might be far from multivariate normal. The numerical performance of the test procedure of Section 2.3 in less extreme cases is illustrated in Section 5.

The bound on the right-hand side of (3.1) decreases at the rate $n^{-1/2}$ as $n$ increases if $q$ remains fixed. If $q$ increases as $n$ increases, the bound is $O(q^2/n^{1/2})$ and converges to zero if $q^4/n \to 0$. In practice, this implies that the left-hand side of (3.1) is likely to be close to zero only if $q^2/n^{1/2}$ is close to zero. The ratio $q^4/n$ is larger than the ratio obtained by several others. Newey and Windmeijer (2009) obtained asymptotic normality with $q^3/n \to 0$. Andrews and Stock (2007b) obtained a similar result for a linear simultaneous equations model. Faster rates of increase of $q$ as a function of $n$ are possible under stronger assumptions. See, for example, Bekker (1994). In contrast to these results, (3.1) is non-asymptotic, holds under weak distributional assumptions, and does not require linearity or simultaneous equations.

To obtain the asymptotic distribution of $T_n(\theta_0)$ under local alternatives, define

(3.2) $\quad \theta_n^* = \theta_0 + n^{-1/2}\kappa$

for some finite $q \times 1$ vector $\kappa$. Let $\{\lambda_j : j = 1,...,q\}$ denote the eigenvalues of $\Sigma$ and $Z_j$ ($j = 1,...,q$) denote the $j$'th component of $Z$. Make

<u>Assumption 4</u>: (i) $\partial g(x,\theta)/\partial \theta$ exists and is a continuous function of $\theta$ for all $\theta$ in a neighborhood of $\theta_0$ and all $X \in \text{supp}(X)$. (ii) $E\sup_{\theta \in \Theta, j,k=1,...,q} |Z_j \partial g(X,\theta)/\partial \theta_k| < \infty$.

Let $\Pi$ denote the orthogonal matrix that diagonalizes $\Sigma$. That is $\Pi\Sigma\Pi' = \Lambda$, where $\Lambda$ is the diagonal matrix whose diagonal elements are the eigenvalues, $\lambda_j$, of $\Sigma$. Let $\gamma_j$ be the $j$'th element of the $q \times 1$ vector

$$\gamma = \Pi\Sigma^{-1/2}E\left[Z\frac{\partial g(X,\theta_0)}{\partial \theta'}\kappa\right].$$

We now have

<u>Theorem 3.2</u>: Let assumptions 1-4 hold. Let $\{\chi_j^2(\gamma_j^2): j = 1,...,q\}$ be independent random variables that are distributed as non-central chi-square with one degree of freedom and non-central parameters $\gamma_j^2$. Under the sequence of local alternatives (3.2)

$$T_n(\theta_0) \to^d \sum_{j=1}^{q} \lambda_j \chi_j^2(\gamma_j^2). \quad \blacksquare$$

Theorem 3.2 implies that the $\alpha$ level test based on $T_n(\theta_0)$ has asymptotic power exceeding $\alpha$ against alternatives whose "distance" from $H_0$ is $O(n^{-1/2})$.

## 4. QUANTILE IV MODELS AND TESTING A PARAMETRIC MODEL AGAINST A NONPARAMETRIC ALTERNATIVE

Section 4.1 treats quantile IV models. Section 4.2 treats tests of model (2.1) and quantile IV models against a nonparametric alternative.

### 4.1 Inference in Quantile IV Models

The quantile model is

(4.1) $\quad Y = g(X, \theta) + U; \quad P(U \leq 0 | Z) = a_Q,$

where $0 < a_Q < 1$. As in model (2.1), $Y$ is the dependent variable, $X$ is a possibly endogenous explanatory variable, and $Z$ is an instrument for $X$. The null hypotheses to be tested are (2.2) and (2.3). However, as is explained in Section 2.2, testing hypothesis (2.3) can be reduced to testing hypothesis (2.2). Therefore, only a test of hypothesis (2.2) is described in this section. Jun (2008) and Andrews and Mikusheva (2016) describe asymptotic tests for quantile IV models that are robust to weak instruments. Other asymptotic tests of (2.2) can be based on any estimation method that yields an estimator of $\theta$ that is asymptotically normally distributed after suitable centering and scaling. Chernozhukov, Hansen, and Jansson (2009) describe an exact finite-sample test of a hypothesis about a parameter in a class of parametric quantile IV models. The test presented in this section is a version of the test presented in Sections 2 and 3. Thus, the same test applies to both mean and quantile IV models and can also be used to test a parametric mean or quantile IV model against a nonparametric alternative.

Let $\{Y_i, X_i, Z_i : i = 1, \dots, n\}$ be an independent random sample from the distribution of $(Y, X, Z)$ in (4.1). Let $Z_{ij}$ $(i = 1, \dots, n; j = 1, \dots, q)$ denote the $j$'th component of $Z_i$. For any $\theta \in \Theta$, define

$$T_{Qn}(\theta) = n^{-1} \sum_{j=1}^{q} \left[ \sum_{i=1}^{n} Z_{ij} W_{Qi}(\theta) \right]^2,$$

where

$$W_{Qi}(\theta) = I[Y_i - g(X_i, \theta) \leq 0] - a_Q.$$

Define

$$W_Q(\theta) = I[Y - g(X, \theta) \leq 0] - a_Q.$$

Denote the covariance matrix of the random vector $ZW(\theta)$ by $\Sigma_Q(\theta)$. Define $\Sigma_Q = \Sigma_Q(\theta_0)$, and let $\hat{\Sigma}_Q$ be the consistent estimator of $\Sigma_Q$ that is defined in the next paragraph. The statistic for testing hypothesis (2.2) is $T_{Qn}(\theta_0)$. Let $\hat{V}_Q$ be a $q \times 1$ random vector that is distributed as $N(0, \hat{\Sigma}_Q)$. Define

(4.2) $\quad \hat{T}_{Qn}(\theta_0) = \hat{V}_Q' \hat{V}_Q.$

Let $\hat{c}_{Q\alpha}(\theta_0)$ denote the $1 - \alpha$ quantile of the distribution of $\hat{T}_{Qn}(\theta_0)$.

The test procedure is:

1. Estimate $\Sigma_Q$ using the estimator $\hat{\Sigma}_Q$ consisting of the $q \times q$ matrix whose $(j, k)$ component is

$$\hat{\Sigma}_{Qjk} = n^{-1}\sum_{i=1}^{n} Z_{ij} Z_{ik} W_{Qi}(\theta_0)^2 - \hat{\mu}_{Qj}\hat{\mu}_{Qk},$$

where

$$\hat{\mu}_{Qj} = n^{-1}\sum_{i=1}^{n} Z_{ij} W_{Qi}(\theta_0).$$

2. Use simulation to compute the distribution of $\hat{T}_{Qn}(\theta_0)$ and the critical value $\hat{c}_{Q\alpha}(\theta_0)$ by repeatedly drawing $\hat{V}_Q$ from the $N(0,\hat{\Sigma}_Q)$ distribution.

3. Reject $H_0$ at the $\alpha$ level if $T_{Qn}(\theta_0) > \hat{c}_{Q\alpha}(\theta_0)$

To obtain a non-asymptotic upper bound on $|P[T_{Qn}(\theta_0) > \hat{c}_{Q\alpha}(\theta_0)] - \alpha|$ make the following assumptions.

<u>Assumption Q1</u>: $\{Y_i, X_i, Z_i : i = 1,...,n\}$ is an independent random sample from the distribution of $(Y,X,Z)$.

<u>Assumption Q2</u>: (i) $\Sigma_Q(\theta)$ is nonsingular for every $\theta \in \Theta$. (ii) Let $\Sigma^{-1}_{Qjk}(\theta)$ denote the $(j,k)$ component of $\Sigma_Q^{-1}(\theta)$. There is a constant $C_{Q\Sigma} < \infty$ such that $|\Sigma^{-1}_{Qjk}(\theta)| \le C_{Q\Sigma}$ for each $j,k = 1,...,q$ and every $\theta \in \Theta$.

Define the $q \times 1$ vectors $\xi_Q = ZW_Q(\theta_0)$ and $\zeta_Q = \Sigma_Q^{-1/2}\xi_Q$. Define the $q \times q$ matrix $\eta_Q = ZZ'W_Q(\theta_0)^2$ Let $\xi_{Qj}$ and $\zeta_{Qj}$ ($j = 1,...,q$) denote the $j$'th components of $\xi_Q$ and $\zeta_Q$, respectively. Let $\eta_{Qjk}$ ($j,k = 1,...,q$) denote the $(j,k)$ component of $\eta_Q$.

<u>Assumption Q3</u>: (i) There is a finite constant $m_3$ such that $E|\zeta_{Qj}|^3 \le m_3$ for every $j = 1,...,q$. (ii) There is a finite constant $\ell_Q \ge \max[\max_j E(\xi_{Qj}^2), \max_{j,k} E(\eta_{Qjk}^2)]$ such that $E|\xi_{Qj}|^r \le \ell_Q^{r-1}r!$ and $E|\eta_{Qjk}|^r \le \ell_Q^{r-1}r!$ for every $r = 3,4,5,...$ and $j,k = 1,...,q$.

For any $t > 0$ define

$$r_Q(t) = \left(\frac{6\ell_Q t}{n}\right)^{1/2}$$

and

$$\tilde{r}_Q(t) = C_{Q\Sigma} q^2 [r_Q(t) + r_Q(t)^2].$$

The following theorem gives the non-asymptotic bound on $|P[T_{Qn}(\theta_0) > \hat{c}_{Q\alpha}(\theta_0)] - \alpha|$.

Theorem 4.1: Let assumptions Q1-Q3 and hypothesis (2.2) hold. If $\max[q\tilde{r}(t), r(t)] < 1$, then

$$(4.3) \quad |P[T_{Qn}(\theta_0) > \hat{c}_{Q\alpha}(\theta_0)] - \alpha| \leq \frac{400 q^{7/4} m_3}{n^{1/2}} + \min \begin{cases} q 2^{q+1} \tilde{r}_Q(t - 2\log q) \\ \frac{1}{\sqrt{2}} \{\tilde{r}_Q(t - 2\log q) - \log[1 - \tilde{r}_Q(t - 2\log q)]\}^{1/2} \end{cases}$$

with probability at least $1 - 4e^{-t}$. ∎

The treatment of the critical value $\hat{c}_{Q\alpha}(\theta_0)$ in (4.3) is the same as that of $\hat{c}_\alpha(\theta_0)$ in (3.1), and the interpretation of the probabilities in Theorem 4.1 is the same as in Theorem 3.1.

The asymptotic distribution of $T_{Qn}(\theta_0)$ under the sequence of local alternative hypotheses (3.2) is given in Theorem 4.2 (iii).

### 4.2 *Testing a Parametric Model against a Nonparametric Alternative*

This section explains how the methods of Sections 2 and 4.1 can be used to carry out a test of a parametric mean or quantile IV model against a nonparametric alternative. As in Sections 2 and 4.1, the method presented in this section provides a finite-sample (non-asymptotic) bound on the difference between the true and nominal probabilities of rejecting a correct null hypothesis. Horowitz (2006) and Horowitz and Lee (2009) describe an asymptotic tests of models (2.1) and (4.1) against nonparametric alternatives. The tests described in this section are non-asymptotic.

Consider, first, model (2.1). Let $G$ be a function that is identified by the relation

(4.4) $\quad E[Y - G(X) | Z] = 0$,

where $Y$, $X$, and $Z$ are as defined in Section 2.1. The null hypothesis, $H_0^{NP}$, tested in this section is

(4.5) $\quad G(x) = g(x, \theta)$

for some $\theta \in \Theta$ and almost every $x \in \text{supp}(X)$, where $g$ is a known function. The alternative hypothesis, $H_1^{NP}$, is that there is no $\theta \in \Theta$ such that (4.5) holds for almost every $x \in \text{supp}(X)$. The sequence of local alternatives used to obtain the asymptotic distribution of the test under $H_1^{NP}$ is

(4.6) $\quad G(X) = g(X, \theta_0) + n^{-1/2} \Delta(X)$,

for some $\theta_0 \in \Theta$, where $\Delta(x)$ a function such that $E|Z_j \Delta(X)| < \infty$. To carry out the test, define $T_n(\theta)$ as in Section 2.1 and $\hat{c}_\alpha(\theta)$ as in Section 2.3 after replacing $\theta_0$ with $\theta$. The test of $H_0^{NP}$ consists of solving the optimization problem

(4.7) $\quad \underset{\theta \in \Theta}{\text{minimize}}: [T_n(\theta) - \hat{c}_\alpha(\theta)]$.

$H_0^{NP}$ is rejected at the $\alpha$ level if the optimal value of the objective function in (4.7) exceeds zero. Theorem 3.1 provides a non-asymptotic upper bound on $|P[T_n(\theta_0) > \hat{c}_\alpha(\theta_0)] - \alpha|$ under $H_0$ and, therefore, on $|P[T_n(\theta) > \hat{c}_\alpha(\theta)] - \alpha|$ for any $\theta \in \Theta$.

Now consider model (4.1). The test of $H_0^{NP}$ for model (4.1) consists of solving the optimization problem

$$\underset{\theta \in \Theta}{\text{minimize}}: [T_{Qn}(\theta) - \hat{c}_{Q\alpha}(\theta)].$$

Theorem 4.1 provides a non-asymptotic upper bound on $|P[T_{Qn}(\theta_0) > \hat{c}_{Q\alpha}(\theta_0)] - \alpha|$ and, therefore, on $|P[T_{Qn}(\theta) > \hat{c}_{Q\alpha}(\theta)] - \alpha|$ for any $\theta \in \Theta$.

We now obtain the asymptotic distributions of $T_n(\theta_0)$ and $T_{Qn}(\theta_0)$ under the nonparametric local alternative (4.6). We also obtain the asymptotic distribution of $T_{nQ}(\theta_0)$ under the parametric local alternative (3.2). Let $f_{U|X,Z}$ denote the probability density of $U$ conditional on $X, Z$ whenever this quantity exists. Make assumption Q5 for model (4.1) and assumption Q6 for models (2.1) and (4.1).

Assumption Q4: (i) There is a neighborhood $\mathcal{N}$ of $u = 0$ such that for all $u \in \mathcal{N}$ and all $(x, z) \in \text{supp}(X, Z)$, $f_{U|X,Z}(u)$ exists, $f_{U|X,Z}(u)$ is a continuous function of $u$, and $|f_{U|X,Z}(u)| \leq M_1$ for all $u$, and $(x, z)$ and some constant $M_1 < \infty$. (ii) $E \sup_{\theta \in \Theta, j,k=1,...,q} |Z_j \partial g(X, \theta)/\partial \theta_k| < \infty$.

Assumption Q5: (i) Alternative hypothesis (4.6) holds. (ii) $E|Z_j \Delta(X)| < \infty$.

Let $\{\lambda_{Qj}: j = 1, 2, ..., q\}$ denote the eigenvalues of $\Sigma_Q$. Let $\Pi_Q$ denote the orthogonal matrix that diagonalizes $\Sigma_Q$. Define $\Pi$ as in Section 3. Let $\tau_j$ be the $j$'th element of the $q \times 1$ vector

$$\tau = \Pi \Sigma^{-1/2} E[Z \Delta(X)].$$

Let $\gamma_{Qj}$ be the $j$'th element of the $q \times 1$ vector

$$\gamma_Q = \Pi_Q \Sigma^{-1/2} E_{XZ}\left[Z \frac{\partial g(X, \theta_0)}{\partial \theta'} \kappa f_{U|X,Z}(0 | X, Z)\right].$$

Let $\tau_{Qj}$ be the $j$'th element of the $q \times 1$ vector

$$\tau_Q = -\Pi_Q \Sigma^{-1/2} E_{XZ}\left[Z \Delta(X) f_{U|X,Z}(0 | X, Z)\right],$$

where $\kappa$ is as in (3.2). We now have

Theorem 4.2: (i) (Model 2.1 with a nonparametric alternative hypothesis). Let assumptions 1, 2, and Q5 hold. Let $\{\chi_j^2(\tau_j^2): j = 1, ..., q\}$ be independent random variables that are distributed as non-central chi-square with one degree of freedom and non-central parameters $\tau_j^2$. Under the sequence of local alternatives (4.6)

$$T_n(\theta_0) \to^d \sum_{j=1}^{q} \lambda_j \chi_j^2(\tau_j^2).$$

(ii) (Model 4.1 with a nonparametric alternative hypothesis). Let assumptions Q1-Q3, Q4(i), and Q5 hold. Let $\{\chi_j^2(\tau_{Qj}^2): j=1,...,q\}$ be independent random variables that are distributed as non-central chi-square with one degree of freedom and non-central parameters $\tau_{Qj}^2$. Under the sequence of local alternatives (4.6)

$$T_{Qn}(\theta_0) \to^d \sum_{j=1}^{q} \lambda_{Qj} \chi_j^2(\tau_{Qj}^2) \ .$$

(iii) (Model 4.1 with a parametric alternative hypothesis). Let assumptions Q1, Q2, and Q4 hold. Let $\{\chi_j^2(\gamma_{Qj}^2): j=1,...,q\}$ be independent random variables that are distributed as non-central chi-square with one degree of freedom and non-central parameters $\gamma_{Qj}^2$. Under the sequence of local alternatives (3.2)

$$T_n(\theta_0) \to^d \sum_{j=1}^{q} \lambda_{Qj} \chi_j^2(\gamma_{Qj}^2). \blacksquare$$

Theorems 3.2 and 4.2 imply that $\alpha$ level tests based on $T_n(\theta_0)$ and $T_{Qn}(\theta_0)$ have asymptotic power exceeding $\alpha$ against parametric and nonparametric alternatives whose "distance" from $H_0$ is $O(n^{-1/2})$.

## 5. MONTE CARLO EXPERIMENTS

This section reports the results of a Monte Carlo investigation of the numerical performance of the test procedure described in Section 2.2. Section 5.1 presents the results of experiments with a correct null hypothesis. Section 5.2 presents results about the power of the test.

### 5.1 *Probability of Rejecting a Correct Null Hypothesis*

The probability of rejecting the correct composite hypothesis (2.3) cannot exceed the probability of rejecting the correct simple hypothesis (2.2) with $\theta_0 = (\vartheta_0', \beta_0')$ for some $\beta_0$ such that $\theta_0$ satisfies (2.1). Therefore, an upper bound on the probability of rejecting a correct simple or composite hypothesis can be obtained by carrying out an experiment with a simple hypothesis. Accordingly, experiments for correct null hypotheses were carried out only for simple hypotheses. When a simple hypothesis is correct,

$$T_n(\theta_0) = n^{-1} \sum_{j=1}^{q} \left[ \sum_{i=1}^{n} Z_{ij} U_i \right]^2 \ .$$

The distribution of $T_n(\theta_0)$ does not depend on the function $g$ or the distribution of $X$, so these are not specified in the designs of the experiments.

Experiments were carried out with sample sizes of $n=100$ and $n=1000$, and with $q=1$, 2, 5, and 10 instruments. The instruments were sampled independently from the $N(0,1)$ distribution. Six distributions of $U$ were used. These are:

1. The uniform distribution: $U \sim U[-2,2]$.

2. A mixture of the $N(0,1)$ and $N(2.5,1)$ distributions centered so that $U$ has mean 0. The mixing probabilities are $p=0.75$ and $p=0.25$, respectively, for the $N(0,1)$ and $N(2.5,1)$ distributions. The resulting mixture distribution is skewed.

3. A mixture of the $N(0,1)$ and $N(4,1)$ distributions centered so that $U$ has mean 0. The mixing probabilities are $p=0.75$ and $p=0.25$, respectively, for the $N(0,1)$ and $N(4,1)$ distributions. The resulting mixture distribution is bimodal.

4. The Laplace distribution..

5. The Student $t$ distribution with 10 degrees of freedom. This distribution does not satisfy assumption 5.

6. The difference between two lognormal distributions.

The nominal rejection probability was 0.05. There were 1000 Monte Carlo replications per experiment.

The results of the experiments are shown in Table 1. The differences between the empirical and nominal probabilities of rejecting $H_0$ are small when $q=1$. The empirical rejection probabilities tend to be below the nominal rejection probability of 0.05 when $n=100$ and $q \geq 2$ or $n=1000$ and $q \geq 5$. This behavior is consistent with Theorem 3.1. When $n$ is fixed and $q$ increases, the difference between the true and nominal rejection probabilities decreases at the rate $q^2/n^{1/2}$. When $n=100$, $q^2/n^{1/2}=0.10$ if $q=1$, but $q^2/n^{1/2}=0.40$ if $q=2$. When $n=1000$, $q^2/n^{1/2}=0.13$ if $q=2$, but $q^2/n^{1/2}=0.79$ if $q=5$. The increases in the differences between the true and nominal rejection probabilities reflect the large increases in the value of $q^2/n^{1/2}$ as $q$ increases from 1 to 2 when $n=100$ and from 2 to 5 when $n=1000$.

5.2 *The Power of the Test*

This section presents Monte Carlo estimates of the power of the $T_n$ test described in Section 2.2. To provide a basis for judging whether the power is high or low, the power of the $T_n$ test is compared with the power of the test of Anderson and Rubin (1949).

In the experiments reported in this section, data were generated from two models, one where $H_0$ is simple and one where it is composite. The model for the simple $H_0$ is

$$Y = \beta_0 X + U$$

$$X = \pi' Z + V$$

$$V = (1 - \rho^2)^{1/2} \varepsilon + \rho U,$$

where $Z \sim N(0, I_q)$; $I_q$ is the $q \times q$ identity matrix; $U$ and $\varepsilon$ have the distributions listed in Section 5.1; $\rho = 0.75$; $\beta_0 = 1.0$ or $\beta_0 = 0.20$, depending on the experiment; and $\pi = ce_q$, where $e_q$ is a $q \times 1$ vector of ones and $c = 0.50$ or $0.25$, depending on the experiment. The instruments are relatively strong when $c = 0.50$ and relatively weak when $c = 0.25$. The null hypothesis is $H_0: \beta = 0$.

The model for the composite $H_0$ is

$$Y = \beta_1 X_1 + \beta_2 X_2 + U$$

$$X_1 = \pi' Z + V$$

$$V = (1 - \rho^2)^{1/2} \varepsilon + \rho U,$$

where $Z \sim N(0, I_q)$; $X_1$ is the endogenous explanatory variable, $X_2$ is exogenous; $X_2$, $U$, and $\varepsilon$ have the distributions listed in Section 5.1; $\rho = 0.75$; $\beta_1 = \beta_2 = 1$ or $\beta_1 = \beta_2 = 0.20$, depending on the experiment; and $\pi = ce_q$, where $c = 0.50$ or $0.25$. The null hypothesis is $H_0: \beta_1 = 0$.

With both models, the sample sizes are $n = 100$ and $n = 1000$, and the numbers of instruments are $q = 1$, 2, 5, and 10. The nominal level of the test is 0.05.

The results of the experiments with the simple $H_0$ are shown in Table 2 for $c = 0.50$ and Table 3 for $c = 0.25$. The results of the experiments with the composite $H_0$ are shown in Table 4 for $c = 0.50$ and Table 5 for $c = 0.25$. In most experiments, the power of the $T_n$ test is similar to the power of the Anderson-Rubin test. This is not surprising because the $T_n$ statistic is a non-Studentized version of the Anderson-Rubin statistic. However, the Anderson-Rubin test is not a substitute for the $T_n$ test. The $T_n$ test applies to nonlinear and quantile models, but the Anderson-Rubin test does not apply to these models.

The power of the $T_n$ test, like that of the Anderson-Rubin test, can be lower than the power of certain other tests if the number of instruments is large. However, the number of instruments is small (often one) in most applications. The power of the $T_n$ test is similar to that of other tests when the number of instruments is small.

# 6. EMPIRICAL APPLICATIONS

This section presents two empirical applications of the $T_n$ test. One consists of testing a hypothesis about a parameter in a finite-dimensional parametric model. The other consists of testing a parametric model against a nonparametric alternative.

## 6.1 *Testing a Hypothesis about a Parameter*

Acemoglu, Johnson, and Robinson (2001) (AJR) estimated models of the effect of institutions on economic performance. We consider the models in columns 1, 2, and 8 of Table IV of AJR. These models have the form

(6.1) $\quad \log(GDP) = \beta_0 + \beta_1 AVPR + \gamma' X + \varepsilon; \quad E(\varepsilon | X, Z)$,

where $GDP$ is a country's GDP per capita. $AVPR$ is an index of protection against expropriation risk. It is the institutional variable of interest and is potentially endogenous. $X$ is a vector of exogenous covariates, $Z$ is an instrument for $AVPR$, and $\varepsilon$ is an unobserved random variable. The instrument $Z$ is the logarithm of European settler mortality in a country. The $\beta$'s and $\gamma$ are constant parameters or vectors. The data consist of observations on 64 countries. The data and instrument are described in AJR. The parameter $\beta_1$ measures the effect of institutions on economic performance.

AJR present two stage least squares estimates of $\beta_1$ for the models in columns 1, 2, and 8 of Table IV and reject the hypothesis that $\beta_1 = 0$ in each column. We use $T_n$ to test the hypothesis $\beta_1 = 0$ in each of the columns. The values of $T_n$ are 50.987, 25.085, and 89864 for columns 1, 2, and 8, respectively. The corresponding 0.05-level critical values are 6.496, 9.461, and 10.845. The $T_n$ test, like AJR, rejects the hypothesis that $\beta_1 = 0$ ($p < 0.05$).

## 6.2 *Testing a Parametric Model against a Nonparametric Alternative*

Blundell, Horowitz, and Parey (2012) (BHP) estimated parametric and non-parametric models of mean gasoline demand conditional on price and income. The parametric model is

(6.1) $\quad \log Q = \beta_0 + \beta_1 \log P + \beta_2 \log Y + \varepsilon; \quad E(\varepsilon | Y, Z) = 0$,

where $Q$ is annual gasoline consumption by a household, $P$ is the potentially endogenous price of gasoline, $Y$ is the household's income, $Z$ is an instrument for $P$, and $\varepsilon$ is an unobserved random variable. The instrument $Z$ is the distance between an oil platform in the Gulf of Mexico and the capital of the household's state. The $\beta$'s are constant parameters. The nonparametric model is

(6.2) $\quad \log Q = g(P, Y) + \varepsilon; \quad E(\varepsilon | Y, Z) = 0$,

where $g$ is an unknown function. The data consist of 4812 observations from the 2001 National Household Travel Survey and are conditioned on a variety of demographic and geographical variables to reduce heterogeneity. BHP provide details about the data and explain the relevance and validity of the instrument.

Figure 2 of BHP shows graphs of the nonparametrically estimated demand function. The function appears nonlinear. We use $T_n$ to test the hypotheses that the parametric model (6.1) is correctly specified against the nonparametric alternative (6.2). The value of $T_n$ is $T_n = 61.21$. The critical value is $\hat{c}_\alpha = 1.36$. The $T_n$ test rejects the hypotheses that the demand function is linear ($p < 0.01$).

## 7. CONCLUSIONS

This paper has presented a non-asymptotic method for carrying out inference in a wide variety of linear and nonlinear models estimated by instrumental variables. "Non-asymptotic" means that the method provides a finite-sample bound on the difference between the true and nominal probabilities of rejecting a correct null hypothesis. The method is a non-Studentized version of the Anderson-Rubin (1949) test but is motivated and analyzed differently. The method is easy to implement and does not require strong distributional assumptions, linearity of the estimated model, or simultaneous equations. Nor does it require knowledge of the strength of the instruments or identification of the parameter about which inference is made. The method can be applied to quantile IV models that may be nonlinear and can be used to test a parametric IV or quantile IV model against a nonparametric alternative. The results presented here hold in finite samples, regardless of the strength of the instruments. The results of Monte Carlo experiments and two empirical applications have illustrated the numerical performance of the method.

## 8. APPENDIX: PROOFS OF THEOREMS

This section presents the proofs of Theorems 3.1, 3.2, 4.1, and 4.2. Assumptions 1-3 and hypothesis (2.2) hold for lemmas 8.1-8.3 and the proof of Theorem 3.1.

<u>Lemma 8.1</u>: Let $\{v_i : i = 1,...,n\}$ be random $q \times 1$ vectors with the $N(0, I_{q \times q})$ distribution. Define

$$\tilde{T}_n(\theta_0) = \left( n^{-1/2} \sum_{i=1}^{n} v_i' \right) \Sigma \left( n^{-1/2} \sum_{i=1}^{n} v_i \right).$$

Then

(8.1) $$\sup_{a \geq 0} | P[T_n(\theta_0) \leq a] - P[\tilde{T}_n(\theta_0) \leq a]| \leq \frac{400q^{7/4}m_3}{n^{1/2}}.$$

Proof: For each $i = 1,...,n$, define

$$\tilde{V}_i = \Sigma^{-1/2}(Z_i U_i).$$

Then $E(\tilde{V}_i) = 0$, $E(\tilde{V}_i \tilde{V}_i') = I_{q \times q}$, and

$$T_n(\theta_0) = \left(n^{-1/2}\sum_{i=1}^{n} \tilde{V}_i\right)' \Sigma \left(n^{-1/2}\sum_{i=1}^{n} \tilde{V}_i\right).$$

For any $a \geq 0$, the set

$$A = \{\tilde{V}_1,...,\tilde{V}_n : T_n(\theta_0) \leq a\}$$

is convex. Therefore, (8.1) follows from Theorem 1.1 of Bentkus (2003). See, also, Corollary 11.1 of Dasgupta (2008). Q.E.D.

Define $r(t)$ as in Theorem 3.1. Define $\omega = \hat{\Sigma} - \Sigma$.

Lemma 8.2: For any $t > 0$ such that

(8.2) $\quad r(t) \leq 1$,

(8.3) $\quad |\omega_{jk}| \leq r(t) + r(t)^2$

uniformly over $j,k = 1,...,q$ with probability at least $1 - 4q^2 e^{-t}$, and

(8.4) $\quad |(\Sigma^{-1}\omega)_{jk}| \leq C_\Sigma q[r(t) + r(t)^2]$

uniformly over $j,k = 1,...,q$ with probability at least $1 - 4q^2 e^{-t}$.

Proof: Define

$$\mu_j = EZ_{1j}[Y_1 - g(X_1, \theta_0)].$$

Then

$$|\omega_{jk}| = n^{-1}\left|\sum_{i=1}^{n}[Z_{ij}Z_{ik}U_i^2 - E(Z_{ij}Z_{ik}U_i^2)] - (\hat{\mu}_j - \mu_j)(\hat{\mu}_k - \mu_k)\right|$$

$$\leq n^{-1}\left|\sum_{i=1}^{n}[Z_{ij}Z_{ik}U_i^2 - E(Z_{ij}Z_{ik}U_i^2)]\right| + |(\hat{\mu}_j - \mu_j)(\hat{\mu}_k - \mu_k)|.$$

Bernstein's inequality gives

$$P\left[n^{-1}\left|\sum_{i=1}^{n}[Z_{ij}Z_{ik}U_i^2 - E(Z_{ij}Z_{ik}U_i^2)]\right| \geq r(t)\right] \leq 2e^{-t}$$

for each $(j,k) = 1,...,q$ and

$$P[|\hat{\mu}_j - \mu_j| \geq r(t)] \leq 2e^{-t}$$

for each $j = 1,...,q$. Therefore,

$$P\left[\max_{j,k} |\omega_{jk}| < r(t) + r(t)^2\right] > 1 - 4q^2 e^{-t},$$

thereby establishing (8.3). In addition,

(8.5) $\quad |(\Sigma^{-1}\omega)_{jk}| \leq C_\Sigma \sum_{\ell=1}^{q} |\omega_{\ell k}|.$

Therefore, inequality (8.4) follows from (8.3) and (8.5). Q.E.D.

Define the random variables $V \sim N(0,\Sigma)$ and, conditional on $\hat{\Sigma}$, $\hat{V} \sim N(0,\hat{\Sigma})$. Also define

$$\Xi_n = \sup_a |P[\tilde{T}_n(\theta_0) \leq a] - P[\hat{T}_n(\theta_0) \leq a]| = \sup_a |P(V'V \leq a) - P(\hat{V}'\hat{V} \leq a)|.$$

<u>Lemma 8.3</u>: Define $\tilde{r}(t)$ as in Theorem 3.1. For any $t > 0$ such that (8.2) holds and $q\tilde{r}(t) < 1$,

$$\Xi_n \leq \min\begin{cases} q2^{q+1}\tilde{r}(t)] \\ \dfrac{1}{\sqrt{2}}\{\tilde{r}(t) - \log[1-\tilde{r}(t)]\}^{1/2} \end{cases}.$$

with probability at least $1 - 2q^2 e^{-t}$.

<u>Proof</u>: Let $TV(P_1,P_2)$ be the total variation distance between distributions $P_1$ and $P_2$. For any set $\mathcal{S} \subset \mathbb{R}^q$ and $q \times 1$ random vector $v$, define and $\mathcal{S}_\Sigma = \{v : \Sigma^{1/2}v \in \mathcal{S}\}$. Then,

$$P(\hat{V} \in \mathcal{S}) - P(V \in \mathcal{S}) = P(\Sigma^{-1/2}\hat{V} \in \mathcal{S}_\Sigma) - P(\Sigma^{-1/2}V \in \mathcal{S}_\Sigma).,$$

By the definition of the total variation distance,

$$\Xi_n \leq \sup_{\mathcal{S}} |P(\Sigma^{-1/2}\hat{V} \in \mathcal{S}_\Sigma) - P(\Sigma^{-1/2}V \in \mathcal{S}_\Sigma)| \leq TV[N(0,I_{q\times q}), N(0,\Sigma^{-1}\hat{\Sigma})],$$

By DasGupta (2008, p. 23),

$$TV[N(0,I_{p\times p}), N(0,\Sigma^{-1}\hat{\Sigma})] \leq \min\begin{cases} q2^{q+1}\|\Sigma^{-1}\hat{\Sigma} - I_{q\times q}\| \\ \dfrac{1}{\sqrt{2}}\left[Tr(\Sigma^{-1}\hat{\Sigma} - I_{q\times q}) - \log\det(\Sigma^{-1}\hat{\Sigma})\right]^{1/2} \end{cases},$$

where for any $q \times q$ matrix $A$,

$$\|A\|^2 = \sum_{j,k=1}^{q} a_{jk}^2.$$

But

$$\Sigma^{-1}\hat{\Sigma} - I_{q\times q} = \Sigma^{-1}(\hat{\Sigma} - \Sigma) = \Sigma^{-1}\omega,$$

$$|(\Sigma^{-1}\omega)_{jk}| \leq \sum_{\ell=1}^{q}|\Sigma_{j\ell}^{-1}\omega_{\ell k}| \leq C_\Sigma \sum_{\ell=1}^{q}|\omega_{\ell k}|,$$

and

$$\left\|\Sigma^{-1}\hat{\Sigma} - I_{q\times q}\right\| \leq C_\Sigma q^{1/2}\left[\sum_{k=1}^{q}\left(\sum_{\ell=1}^{q}|\omega_{\ell k}|\right)^2\right]^{1/2} \leq C_\Sigma q^2 \max_{\ell,k}|\omega_{\ell k}|$$

By Lemma 7.2

$$P\left[\max_{j,k}|\omega_{jk}| < r(t) + r(t)^2\right] > 1 - 4q^2 e^{-t}.$$

Therefore,

$$q2^{q+1}\left\|\Sigma^{-1}\hat{\Sigma} - I_{q\times q}\right\| \leq q2^{q+1}\tilde{r}(t)$$

with probability exceeding $1 - 4q^2 e^{-t}$.

Now consider

$$\frac{1}{\sqrt{2}}\left[Tr(\Sigma^{-1}\hat{\Sigma} - I_{q\times q}) - \log\det(\Sigma^{-1}\hat{\Sigma})\right]^{1/2}.$$

We have

$$Tr(\Sigma^{-1}\hat{\Sigma} - I_{q\times q}) = Tr(\Sigma^{-1}\omega).$$

But

$$(\Sigma^{-1}\omega)_{jj} \leq C_\Sigma q[r(t) + r(t)^2]$$

with probability exceeding $1 - 4q^2 e^{-t}$. Therefore

$$Tr(\Sigma^{-1}\omega) \leq \tilde{r}(t),$$

and

$$\frac{1}{\sqrt{2}}\left[Tr(\Sigma^{-1}\hat{\Sigma} - I_{q\times q}) - \log\det(\Sigma^{-1}\hat{\Sigma})\right]^{1/2} \leq \frac{1}{\sqrt{2}}\left[\tilde{r}(t) - \log\det(\Sigma^{-1}\hat{\Sigma})\right]^{1/2}$$

with probability exceeding $1 - 4q^2 e^{-t}$.

In addition,

$$\log\det(\Sigma^{-1}\hat{\Sigma}) = \log\det(I_{q\times q} + \Sigma^{-1}\omega).$$

Let $\tilde{r}(t) < 1$. By Corollary 1 of Brent, Osborne, and Smith (2015)

$$\det(I_{q\times q} + \Sigma^{-1}\omega) \geq 1 - \tilde{r}(t)$$

and

$$\log\det(I_{q\times q} + \Sigma^{-1}\omega) \geq \log[1 - \tilde{r}(t)]$$

with probability exceeding $1-2q^2e^{-t}$. Therefore,

$$\frac{1}{\sqrt{2}}\left[Tr(\Sigma^{-1}\hat{\Sigma} - I_{p\times p}) - \log\det(\Sigma^{-1}\hat{\Sigma})\right]^{1/2} \le \frac{1}{\sqrt{2}}\{\tilde{r}(t) - \log[1-\tilde{r}(t)]\}^{1/2}$$

and

$$\Xi_n \le \min\begin{cases} C_\Sigma q^3 2^{q+1}\tilde{r}(t)] \\ \frac{1}{\sqrt{2}}\{\tilde{r}(t) - \log[1-\tilde{r}(t)]\}^{1/2}\end{cases}$$

with probability at least $1-4q^2e^{-t}$. Q.E.D.

<u>Proof of Theorem 3.1</u>: By the triangle inequality

$$\sup_{a\ge 0}|P[T_n(\theta_0)\le a] - P[\hat{T}_n(\theta_0)\le a]|$$

$$= \sup_{a\ge 0}|P[T_n(\theta_0)\le a] - P[\tilde{T}_n(\theta_0)\le a] + P[\tilde{T}_n(\theta_0)\le a] - P[\hat{T}_n(\theta_0)\le a]|$$

$$\le \sup_{a\ge 0}\{|P[T_n(\theta_0)\le a] - P[\tilde{T}_n(\theta_0)\le a]| + |P[\tilde{T}_n(\theta_0)\le a] - P[\hat{T}_n(\theta_0)\le a]|\}\ .$$

$$\le \sup_{a\ge 0}|P[T_n(\theta_0)\le a] - P[\tilde{T}_n(\theta_0)\le a]| + \sup_{a\ge 0}|P[\tilde{T}_n(\theta_0)\le a] - P[\hat{T}_n(\theta_0)\le a]|$$

$$\le \sup_{a\ge 0}|P[T_n(\theta_0)\le a] - P[\tilde{T}_n(\theta_0)\le a]| + \Xi_n.$$

Now combine lemmas 7.1 and 7.3. Q.E.D.

<u>Proof of Theorem 3.2</u>: Let $\mathbf{Z}_i$ be the $q\times 1$ vector whose $j$'th component is $Z_{ij}$. A Taylor series expansion yields

$$n^{-1/2}\sum_{i=1}^n \mathbf{Z}_i[Y_i - g(X_i,\theta_0)] = n^{-1/2}\sum_{i=1}^n \mathbf{Z}_i U_i + n^{-1}\sum_{i=1}^n \mathbf{Z}_i\frac{\partial g(X_i,\tilde{\theta})}{\partial\theta'}\kappa,$$

where $\tilde{\theta}$ is between $\theta_n^*$ and $\theta_0$. It follows from the multivariate generalization of the Lindeberg-Levy theorem and Theorem 2 of Jennrich (1969) that

$$n^{-1/2}\sum_{i=1}^n \mathbf{Z}_i[Y_i - g(X,\theta_0)] \to^d \xi + E\left[Z\frac{\partial g(X,\theta_0)}{\partial\theta'}\right]\kappa,$$

where $\xi \sim N(0,\Sigma)$. As in Section 3, let $\Pi$ denote the orthogonal matrix that diagonalizes $\Sigma$. That is $\Pi\Sigma\Pi' = \Lambda$, where $\Lambda$ is the diagonal matrix whose diagonal elements are the eigenvalues, $\lambda_j$, of $\Sigma$. Define

$$\tilde{\gamma} = \Sigma^{-1/2} E\left[Z\frac{\partial g(X,\theta_0)}{\partial \theta'}\right]\kappa.$$

Then

$$T_n(\theta_0) \to^d (\xi + \Sigma^{1/2}\tilde{\gamma})'(\xi + \Sigma^{1/2}\tilde{\gamma})$$

$$= (\Sigma^{-1/2}\xi + \tilde{\gamma})'\Sigma(\Sigma^{-1/2}\xi + \tilde{\gamma})$$

$$= \{\Pi(\Sigma^{-1/2}\xi + \tilde{\gamma})\}'\Lambda\{\Pi(\Sigma^{-1/2}\xi + \tilde{\gamma})\}.$$

The theorem now follows from the properties of quadratic forms of normal random variables. Q.E.D.

Proof of Theorem 4.1: Replace $ZU$ with $ZW(\theta_0)$ in lemmas 7.1-7.3. Then proceed as in the proof of Theorem 3.1. Q.E.D.

Let $F_{XZ}$ denote the distribution function of $(X,Z)$ and $F_{X|Z}$ denote the conditional distribution function of $X$ given $Z$.

Proof of Theorem 4.2:

Part (i): Part (i) follows from the multivariate generalization of the Lindeberg-Levy central limit theorem and the definition of $\tau$.

Part (ii): Arguments like those used in the proof of Theorem 3.2 show that

$$n^{-1/2}\sum_{i=1}^{n} Z_{ij}\{I[Y_i - g(X_i,\theta_0) \leq 0] - a_Q\} \to^d \xi_Q + \tilde{\tau}_{Qj},$$

where $\xi_Q \sim N(0,\Sigma_Q)$ and

$$\tilde{\tau}_{Qj} = \lim_{n\to\infty} n^{1/2} E\left(Z_{1j}\{I[Y - g(X,\theta_0) \leq 0] - a_Q\}\right).$$

Under alternative hypothesis (4.6),

$$\tilde{\tau}_{Qj} = \lim_{n\to\infty} n^{1/2} E Z_{1j}\{P[U \leq n^{-1/2}\Delta(X) | Z_1] - a_Q\}.$$

Now

$$P[U \leq n^{-1/2}\Delta(X) | Z_1] = \int P[U \leq n^{-1/2}\Delta(x) | X = x, Z_1] dF_{X|Z}(x|Z_1)$$

$$= \int F_{U|X,Z}[-n^{-1/2}\Delta(x) | X = x, Z_1] dF_{X|Z}(x|Z_1).$$

By a Taylor series expansion

$$F_{U|X,Z}[-n^{-1/2}\Delta(x) | X = x, Z_1] = F_{U|X,Z}(0 | X = x, Z_1) - n^{-1/2} f_{U|X,Z}(\tilde{u} | X = x, Z_1)\Delta(x),$$

where $\tilde{u}$ is between 0 and $n^{-1/2}\Delta(x)$. Therefore,

$$P[U \leq n^{-1/2}\Delta(X)|Z_1]$$

$$= \int F_{U|X,Z}(0|X=x,Z_1)dF_{X|Z}(x|Z_1) - n^{-1/2}\int f_{U|X,Z}(\tilde{u}|X=x,Z_1)\Delta(x)dF_{X|Z}(x|Z_1)$$

$$= a_Q - n^{-1/2}\int f_{U|X,Z}(\tilde{u}|X=x,Z_1)\Delta(x)dF_{X|Z}(x|Z_1).$$

In addition,

$$\int f_{U|X,Z}(\tilde{u}|X=x,Z_1)\Delta(x)dF_{X|Z}(x|Z_1) \to \int f_{U|X,Z}(0|X=x,Z_1)\Delta(x)dF_{X|Z}(x|Z_1)$$

as $n \to \infty$. Therefore,

$$n^{1/2}\{P[U \leq n^{-1/2}\Delta(X)|Z_1] - a_Q\} \to -\int f_{U|X,Z}(0|X=x,Z_1)\Delta(x)dF_{X|Z}(x|Z_1),$$

and

$$\tilde{\tau}_{Qj} \to -\int z_j \Delta(x) f_{U|X,Z}(0|X=x,Z_1=z)dF_{X|Z}(x|z)dF_Z(z)$$

$$= -\int z_j \Delta(x) f_{U|X,Z}(0|X=x,Z_1=z)dF_{XZ}(x,z)$$

$$= -E_{XZ}[Z_j \Delta(X) f_{U|X,Z}(0|X,Z)].$$

Part (ii) now follows from arguments like those used in the proof of Theorem 3.2.

Part (iii): Under local alternative hypothesis (3.2),

$$g(x,\theta_n) = g(x,\theta_0) + n^{-1/2}\frac{\partial g(x,\theta_0)}{\partial \theta'}\kappa + o(n^{-1/2}).$$

Therefore, the arguments made for local alternative hypothesis (4.6) apply to local alternative hypothesis (3.2) after replacing $\Delta(x)$ with $[\partial g(x,\theta_0)/\partial \theta']\kappa$. It follows that under local alternative hypothesis (3.2)

$$T_n(\theta_0) \to^d \sum_{j=1}^{q} \lambda_{Qj}\chi_j^2(\gamma_{Qj}^2).$$

This proves Theorem 3.2(iii). Q.E.D.

**Table 1: Empirical Probabilities of Rejecting Correct Null Hypotheses at the Nominal 0.05 Level**

| Distr. | $n$ | $q=1$ | $q=2$ | $q=5$ | $q=10$ |
|---|---|---|---|---|---|
| Uniform | 100 | 0.046 | 0.053 | 0.041 | 0.025 |
|  | 1000 | 0.049 | 0.052 | 0.050 | 0.062 |
| Skewed | 100 | 0.053 | 0.039 | 0.036 | 0.030 |
|  | 1000 | 0.050 | 0.049 | 0.033 | 0.037 |
| Bimodal | 100 | 0.052 | 0.035 | 0.041 | 0.035 |
|  | 1000 | 0.056 | 0.044 | 0.034 | 0.038 |
| Laplace | 100 | 0.043 | 0.032 | 0.031 | 0.013 |
|  | 1000 | 0.041 | 0.049 | 0.044 | 0.043 |
| $t(10)$ | 100 | 0.052 | 0.036 | 0.029 | 0.013 |
|  | 1000 | 0.048 | 0.033 | 0.035 | 0.046 |
| Diff. betw. Lognormals | 100 | 0.041 | 0.027 | 0.016 | 0.010 |
|  | 1000 | 0.053 | 0.062 | 0.035 | 0.031 |

**Table 2: Powers of the $T_n$ and Anderson-Rubin Tests of a Simple Null Hypothesis at the Nominal 0.05 Level**

| Distr. | $n$ | $\beta_0$ | $c$ | $q=1$ $T_n$ | $q=1$ AR | $q=2$ $T_n$ | $q=2$ AR | $q=5$ $T_n$ | $q=5$ AR | $q=10$ $T_n$ | $q=10$ AR |
|---|---|---|---|---|---|---|---|---|---|---|---|
| Uniform | 100 | 1.0 | 0.50 | 0.642 | 0.649 | 0.817 | 0.837 | 0.965 | 0.981 | 0.994 | 0.999 |
|  | 1000 | 0.20 | 0.50 | 0.635 | 0.632 | 0.848 | 0.851 | 0.994 | 0.995 | 1.00 | 1.00 |
| Skewed | 100 | 1.0 | 0.50 | 0.439 | 0.454 | 0.581 | 0.617 | 0.827 | 0.884 | 0.944 | 0.978 |
|  | 1000 | 0.20 | 0.50 | 0.436 | 0.433 | 0.655 | 0.659 | 0.924 | 0.920 | 0.989 | 0.990 |
| Bimodal | 100 | 1.0 | 0.50 | 0.270 | 0.280 | 0.366 | 0.377 | 0.561 | 0.619 | 0.712 | 0.829 |
|  | 1000 | 0.20 | 0.50 | 0.269 | 0.271 | 0.417 | 0.420 | 0.643 | 0.658 | 0.849 | 0.854 |
| Laplace | 100 | 1.0 | 0.50 | 0.510 | 0.502 | 0.654 | 0.683 | 0.842 | 0.903 | 0.946 | 0.987 |
|  | 1000 | 0.20 | 0.50 | 0.486 | 0.483 | 0.663 | 0.667 | 0.923 | 0.938 | 0.998 | 0.999 |
| $t(10)$ | 100 | 1.0 | 0.50 | 0.481 | 0.486 | 0.642 | 0.665 | 0.826 | 0.888 | 0.911 | 0.975 |
|  | 1000 | 0.20 | 0.50 | 0.487 | 0.488 | 0.663 | 0.664 | 0.922 | 0.925 | 0.992 | 0.998 |
| Diff. betw. Lognormals | 100 | 1.0 | 0.50 | 0.142 | 0.135 | 0.216 | 0.221 | 0.248 | 0.337 | 0.304 | 0.481 |
|  | 1000 | 0.20 | 0.50 | 0.143 | 0.137 | 0.191 | 0.181 | 0.310 | 0.334 | 0.388 | 0.449 |

**Table 3: Powers of the $T_n$ and Anderson-Rubin Tests of a Simple Null Hypothesis at the Nominal 0.05 Level**

| Distr. | $n$ | $\beta_0$ | $c$ | $q=1$ $T_n$ | $q=1$ AR | $q=2$ $T_n$ | $q=2$ AR | $q=5$ $T_n$ | $q=5$ AR | $q=10$ $T_n$ | $q=10$ AR |
|---|---|---|---|---|---|---|---|---|---|---|---|
| Uniform | 100 | 1.0 | 0.25 | 0.226 | 0.235 | 0.309 | 0.317 | 0.420 | 0.482 | 0.558 | 0.661 |
|  | 1000 | 0.20 | 0.25 | 0.303 | 0.298 | 0.412 | 0.418 | 0.674 | 0.685 | 0.878 | 0.890 |
| Skewed | 100 | 1.0 | 0.25 | 0.144 | 0.150 | 0.190 | 0.201 | 0.289 | 0.332 | 0.332 | 0.444 |
|  | 1000 | 0.20 | 0.25 | 0.194 | 0.202 | 0.308 | 0.305 | 0.473 | 0.475 | 0.648 | 0.657 |
| Bimodal | 100 | 1.0 | 0.25 | 0.105 | 0.101 | 0.125 | 0.129 | 0.176 | 0.204 | 0.173 | 0.251 |
|  | 1000 | 0.20 | 0.25 | 0.127 | 0.125 | 0.184 | 0.185 | 0.269 | 0.265 | 0.366 | 0.364 |
| Laplace | 100 | 1.0 | 0.25 | 0.195 | 0.175 | 0.234 | 0.231 | 0.278 | 0.326 | 0.330 | 0.466 |
|  | 1000 | 0.20 | 0.25 | 0.210 | 0.207 | 0.296 | 0.286 | 0.462 | 0.467 | 0.679 | 0.695 |
| $t(10)$ | 100 | 1.0 | 0.25 | 0.163 | 0.159 | 0.206 | 0.210 | 0.303 | 0.361 | 0.345 | 0.497 |
|  | 1000 | 0.20 | 0.25 | 0.203 | 0.204 | 0.275 | 0.274 | 0.436 | 0.453 | 0.673 | 0.703 |
| Diff. betw. Lognormals | 100 | 1.0 | 0.25 | 0.061 | 0.064 | 0.085 | 0.087 | 0.080 | 0.111 | 0.058 | 0.157 |
|  | 1000 | 0.20 | 0.25 | 0.092 | 0.089 | 0.086 | 0.093 | 0.121 | 0.129 | 0.139 | 0.178 |

**Table 4: Powers of the $T_n$ and Anderson-Rubin Tests of a Composite Null Hypothesis at the Nominal 0.05 Level**

| Distr. | $n$ | $\beta_1, \beta_2$ | $c$ | $q=1$ $T_n$ | $q=1$ AR | $q=2$ $T_n$ | $q=2$ AR | $q=5$ $T_n$ | $q=5$ AR | $q=10$ $T_n$ | $q=10$ AR |
|---|---|---|---|---|---|---|---|---|---|---|---|
| Uniform | 100 | 1.0 | 0.50 | 0.427 | 0.387 | 0.643 | 0.590 | 0.907 | 0.899 | 0.981 | 0.997 |
|  | 1000 | 1.0 | 0.50 | 1.0 | 1.0 | 1.0 | 1.0 | 1.0 | 1.0 | 1.0 | 1.0 |
|  | 1000 | 0.20 | 0.50 | 0.432 | 0.396 | 0.661 | 0.630 | 0.923 | 0.917 | 1.0 | 0.999 |
| Skewed | 100 | 1.0 | 0.50 | 0.293 | 0.258 | 0.378 | 0.345 | 0.714 | 0.693 | 0.882 | 0.903 |
|  | 1000 | 1.0 | 0.50 | 0.998 | 0.998 | 1.0 | 1.0 | 1.0 | 1.0 | 1.0 | 1.0 |
|  | 1000 | 0.20 | 0.50 | 0.320 | 0.276 | 0.395 | 0.362 | 0.740 | 0.721 | 0.920 | 0.918 |
| Bimodal | 100 | 1.0 | 0.50 | 0.127 | 0.093 | 0.215 | 0.156 | 0.398 | 0.321 | 0.571 | 0.558 |
|  | 1000 | 1.0 | 0.50 | 0.912 | 0.943 | 0.999 | 0.998 | 1.0 | 1.0 | 1.0 | 1.0 |
|  | 1000 | 0.20 | 0.50 | 0.136 | 0.122 | 0.223 | 0.171 | 0.432 | 0.338 | 0.615 | 0.610 |
| Laplace | 100 | 1.0 | 0.50 | 0.307 | 0.252 | 0.457 | 0.381 | 0.716 | 0.707 | 0.886 | 0.928 |
|  | 1000 | 1.0 | 0.50 | 0.999 | 0.999 | 1.0 | 1.0 | 1.0 | 1.0 | 1.0 | 1.0 |
|  | 1000 | 0.20 | 0.50 | 0.342 | 0.309 | 0.461 | 0.420 | 0.742 | 0.728 | 0.910 | 0.936 |
| $t(10)$ | 100 | 1.0 | 0.50 | 0.308 | 0.224 | 0.464 | 0.410 | 0.687 | 0.674 | 0.860 | 0.913 |
|  | 1000 | 1.0 | 0.50 | 0.998 | 0.999 | 1.0 | 1.0 | 1.0 | 1.0 | 1.0 | 1.0 |
|  | 1000 | 0.20 | 0.50 | 0.328 | 0.288 | 0.477 | 0.430 | 0.710 | 0.688 | 0.881 | 0.910 |
| Diff. betw. Lognormals | 100 | 1.0 | 0.50 | 0.075 | 0.053 | 0.097 | 0.073 | 0.171 | 0.143 | 0.189 | 0.210 |
|  | 1000 | 1.0 | 0.50 | 0.543 | 0.648 | 0.876 | 0.819 | 0.989 | 0.985 | 1.0 | 1.0 |
|  | 1000 | 0.20 | 0.50 | 0.083 | 0.077 | 0.099 | 0.084 | 0.182 | 0.166 | 0.220 | 0.213 |

**Table 5: Powers of the $T_n$ and Anderson-Rubin Tests of a Composite Null Hypothesis at the Nominal 0.05 Level**

| Distr. | $n$ | $\beta_1, \beta_2$ | $\pi$ | $q=1$ $T_n$ | $q=1$ AR | $q=2$ $T_n$ | $q=2$ AR | $q=5$ $T_n$ | $q=5$ AR | $q=10$ $T_n$ | $q=10$ AR |
|---|---|---|---|---|---|---|---|---|---|---|---|
| Uniform | 100 | 1.0 | 0.25 | 0.116 | 0.076 | 0.138 | 0.120 | 0.270 | 0.263 | 0.431 | 0.391 |
| | 1000 | 1.0 | 0.25 | 0.896 | 0.846 | 0.996 | 0.981 | 1.0 | 1.0 | 1.0 | 1.0 |
| | 1000 | 0.20 | 0.25 | 0.137 | 0.062 | 0.229 | 0.240 | 0.483 | 0.462 | 0.791 | 0.685 |
| Skewed | 100 | 1.0 | 0.25 | 0.073 | 0.043 | 0.110 | 0.070 | 0.157 | 0.116 | 0.243 | 0.201 |
| | 1000 | 1.0 | 0.25 | 0.688 | 0.605 | 0.905 | 0.856 | 0.997 | 0.992 | 1.0 | 1.0 |
| | 1000 | 0.20 | 0.25 | 0.082 | 0.035 | 0.125 | 0.090 | 0.278 | 0.250 | 0.524 | 0.488 |
| Bimodal | 100 | 1.0 | 0.25 | 0.050 | 0.029 | 0.042 | 0.028 | 0.069 | 0.039 | 0.097 | 0.083 |
| | 1000 | 1.0 | 0.25 | 0.403 | 0.326 | 0.642 | 0.525 | 0.916 | 0.846 | 0.993 | 0.984 |
| | 1000 | 0.20 | 0.25 | 0.060 | 0.042 | 0.076 | 0.058 | 0.126 | 0.088 | 0.231 | 0.185 |
| Laplace | 100 | 1.0 | 0.25 | 0.070 | 0.046 | 0.101 | 0.073 | 0.151 | 0.125 | 0.213 | 0.204 |
| | 1000 | 1.0 | 0.25 | 0.708 | 0.621 | 0.921 | 0.872 | 1.0 | 0.999 | 1.0 | 1.0 |
| | 1000 | 0.20 | 0.25 | 0.080 | 0.042 | 0.159 | 0.090 | 0.320 | 0.294 | 0.500 | 0.465 |
| $t(10)$ | 100 | 1.0 | 0.25 | 0.078 | 0.054 | 0.096 | 0.066 | 0.159 | 0.110 | 0.215 | 0.200 |
| | 1000 | 1.0 | 0.25 | 0.693 | 0.613 | 0.920 | 0.847 | 0.996 | 0.992 | 1.0 | 1.0 |
| | 1000 | 0.20 | 0.25 | 0.104 | 0.042 | 0.156 | 0.110 | 0.273 | 0.255 | 0.533 | 0.488 |
| Diff. betw. Lognormals | 100 | 1.0 | 0.25 | 0.040 | 0.026 | 0.045 | 0.027 | 0.035 | 0.024 | 0.055 | 0.054 |
| | 1000 | 1.0 | 0.25 | 0.153 | 0.115 | 0.261 | 0.173 | 0.492 | 0.382 | 0.734 | 0.635 |
| | 1000 | 0.20 | 0.25 | 0.039 | 0.031 | 0.040 | 0.028 | 0.046 | 0.022 | 0.074 | 0.048 |